\PassOptionsToPackage{authoryear,round}{natbib}
\documentclass[preprint,10pt]{elsarticle}

\usepackage{amsmath,amsfonts,amssymb}
\usepackage{caption,tabularx,booktabs}
\usepackage{multicol}
\usepackage{tikz}
\usepackage{xcolor}
\usepackage{hyperref}
\usetikzlibrary {arrows.meta}
\usepackage[para,online,flushleft]{threeparttable}
\usepackage{bm}
\usepackage{mathtools}
\DeclareMathOperator*{\argmin}{arg\,min}
\usepackage{siunitx}

\usepackage[margin=2cm]{geometry}

\usepackage[authoryear]{natbib}

\begin{document}


\begin{frontmatter}
\title{Parameter estimation for land-surface models using Neural Physics}

\author[1]{Ruiyue Huang}
\author[2,3]{Claire E. Heaney}
\author[1]{Maarten van Reeuwijk\corref{cor1}}
\ead{m.vanreeuwijk@imperial.ac.uk}

\cortext[cor1]{Corresponding author}

\affiliation[1]{organization={Department of Civil and Environmental Engineering, Imperial College London}, addressline={SW7 2AZ London}, country={UK}}

\affiliation[2]{organization={Department of Earth Science and Engineering, Imperial College London}, addressline={SW7 2AZ London}, country={UK}}

\affiliation[3]{organization={Imperial-X, Imperial College London}, addressline={W12 7SL London},
 country={UK}}

\begin{abstract}
We propose a novel inverse-modelling approach that estimates the parameters of a simple land-surface model (LSM) by assimilating data into a differentiable, physics-based forward model formulated using convolutional operations. The governing equations are expressed within the Neural Physics framework, allowing direct gradient-based optimisation of time-dependent parameters without the need to derive and maintain adjoint formulations. The model parameters are estimated by minimising the mismatch between model predictions and synthetic or observational data. Although differentiability is enabled through machine-learning libraries, the forward model itself remains entirely physics-based and neither the forward model nor the parameter estimation procedure involve training.

To evaluate the approach, we first generate synthetic observations of soil temperature by running the forward model with known parameter values and subsequently treat these parameters as unknown in an inverse problem. We show that observations of soil temperature at a single depth are insufficient to reliably constrain the model parameters. Using observations at two depths, however, does yield reliable parameter estimates, although the individual contributions of latent and sensible heat fluxes cannot be distinguished.

We also apply the approach to urban flux tower data from Phoenix, United States, and show that the thermal conductivity, volumetric heat capacity and the combined sensible-latent heat transfer coefficient can be reliably estimated whilst using an observed value for the effective surface albedo. The resulting model accurately predicts the outgoing longwave radiation, conductive soil fluxes and the combined sensible-latent heat fluxes, demonstrating that the Neural Physics framework can be used to accurately determine the parameters of the particular LSM used here. This model is intentionally simple and does not include, for example, a subsurface moisture model. This simplicity facilitates an exploration of parameter identifiability, confounding and equifinality.
\end{abstract}

\end{frontmatter}

\section{Introduction}  The exchange of momentum and thermal energy between the land surface and the atmosphere is of great importance to atmospheric processes and thermal comfort \citep{Oke2017urban}. At present, such processes are typically modelled with land-surface models (LSMs) using meteorological forcing data and various parameters associated with the site. Many LSMs have been developed over the years, each differing in how they handle the urban surface, and in their modelling of various morphological, vegetation, hydrological and anthropogenic processes. Modern LSMs are capable of closely replicating the heterogeneous surface structure of cities \citep{leelee2020MUSE, krayenhoff2007microscale}, and capturing the effects of various urban processes, such as urban heat storage and vegetation dynamics, using various soil, hydrological or vegetation models \citep{Lipson2017heatstorage,Arsenault2018NoahMPDynamicVegetation}. 

However, larger complexity of LSMs does not always translate to better predictions \citep{Grimmond2010PILPSPhase1,Grimmond2011PILPSPhase2,Lipson2024UP}. The Urban-PLUMBER project \citep{Lipson2024UP} evaluated the performance of 30 LSMs in estimating surface fluxes at a suburban site in Melbourne, Australia.  
The study found that models with highly complex urban schemes often do not perform as well as models with simpler urban schemes, as their performances are penalised by having relatively simple representation of hydrological/vegetation models. Most complex models were also unable to make efficient use of site-specific information.   

Besides capturing the complexity of land-surface interactions, another key challenge in modelling urban surface fluxes lies in estimating (or calibrating) the relevant input parameters \citep{Lipson2017heatstorage,Chaney2016Noah}. LSMs typically have different sensitivities to changes in different parameters, making it difficult to perform data assimilation or improve model accuracy by manually updating parameter values, as a slight alteration can lead to large and unexpected changes in the final output \citep{Massoud2019parameter}. \citet{Grimmond2011PILPSPhase2} found that poor estimation of parameter values can result in a significant decline in performance of LSMs. It is hence critical to find a reliable method of parameter estimation.   

\citet{Raoult2024paraestimation} presented an overview of the popular techniques for estimating the parameters of LSMs, which include statistical methods and inverse modelling. Statistical methods involve estimating the probability distributions of model parameters based on prior knowledge and observed data. Some examples include the Markov Chain Monte Carlo (MCMC) method \citep{hastings1970}, which draws samples from posterior distributions to determine the likely values of parameters. However, statistical methods can be computationally intensive, and therefore have been mainly used in computationally inexpensive LSMs such as Simplified PnET (SIPNET) ecosystem model \citep{Fer2018}, or to estimate isolated processes (e.g. plant respiration \citep{Jones2024}). Inverse modelling employs optimisation algorithms, such as the gradient-descent method or genetic algorithms, to determine the optimal set of parameters that minimises the error between the model output and observational data. Such minimisation algorithms have been shown to have better overall optimisation efficiency than statistical methods \citep{Raoult2024paraestimation}. However, inverse modelling algorithms, especially those based on the gradient-descent method, are prone to become stuck in local minima, leading to non-unique solutions that are dependent on the initial conditions \citep{Kuppel2014optimisation,Raoult2016optimisation}. 

A number of researchers have begun exploring the coupling of machine-learning models with differentiable physics-based models to construct differentiable hybrid models~\citep{Shen2023}. In these approaches, neural networks are embedded within a physics-based framework, which is often trained end-to-end, to infer unknown parameters or represent unknown physical relationships. Several studies focus on learning physical processes or closure relationships. For example, \citet{Fang2024} proposed a hybrid model to learn the impact of soil water stress on photosynthesis and evapotranspiration, outperforming traditional terrestrial biosphere models that use empirical soil water stress parametrisations. In a similar vein, \citet{ElGhawi2023} represent leaf surface and aerodynamic resistances using two neural networks within an evapotranspiration model, while \citet{Norouzi2026} use neural networks to separate adsorption and capillary effects on the soil water retention curve. Other work embeds neural networks within land-surface models, learning a snow process representation~\citep{Deck2026} and constitutive relationships for frozen soil~\citep{Ren2026}. Hybrid models have also been applied to inverse problems and parameter estimation. For example, \citet{Ouyang2025} re-implemented the Xin'anjiang (XAJ) hydrological model in PyTorch representing the discretised equations for run-off and evapotranspiration by neural networks. An inverse problem is solved to learn the weights of an LSTM model which predicts the 15~parameters of the XAJ model. With limited data, their model gave better streamflow predictions than manually calibrating XAJ. A further example is JAX-CanVeg~\citep{Jiang2025}, in which a multi-layer land-surface canopy model is augmented with a neural-network--based closure for leaf relative humidity, improving predictions of latent heat and net ecosystem exchange across four flux tower sites. These studies demonstrate the potential of differentiable hybrid models to improve the performance and calibration of LSMs. However, challenges still remain for these differentiable hybrid methods, including sensitivity to training data; tendency of predictions to become physically inconsistent; and computational cost, arising when physics-based models require many forward simulations for parameter optimisation, particularly in end-to-end differentiable settings~\citep{Fang2024,Dukes2026}. 

A related class of differentiable but fully physics-based models has also been developed~\citep{Zhu2021, Bezgin2023, Chen2024}. One such training-free approach, called Neural Physics, capitalises on the equivalence between discretisation stencils and convolutional operators, allowing partial differential equations and their solvers to be expressed as the convolutional layers of neural networks with analytically determined weights. While differentiability can also be achieved using general-purpose automatic differentiation frameworks (e.g.~Julia, or JAX and NumPy-based implementations), Neural Physics provides an alternative to these by formulating numerical discretisations of physical laws using machine-learning primitives such as convolutional operations. The motivation for using Neural Physics in this work extends beyond differentiability. Among its principal advantages are its established and versatile framework, differentiability, reduced implementation complexity and support for a broad range of numerical methods and hardware architectures. The forward model has been applied to a wide range of problems including single-phase flow benchmarks~\citep{Chen2024}, multiphase flow~\citep{Chen2024multiphase}, shallow water dynamics~\citep{Chen2024SWE} and applications in nuclear engineering~\citep{Phillips2023}. It is also platform agnostic, being able to execute on CPUs, GPUs and AI accelerators~\citep{CS3,GraphCore} without modification. The end-to-end differentiability of Neural Physics enables the computation of sensitivities of objective functions with respect to model parameters, allowing standard optimisation tools from machine-learning libraries to be used for parameter estimation. Furthermore, the use of convolutional operations and other functions from machine-learning libraries abstracts away much of the implementation complexity associated with accelerator portability and adjoint development. Neural Physics and the associated NN4PDEs code also supports a broader class of numerical methods, including higher-order discretisations~\citep{Phillips2024}; finite volume~\citep{Xue2026}, finite element and finite difference discretisations~\citep{Chen2024} and multigrid solvers~\citep{Phillips2023}. The latter can be exactly represented as a U-Net architectures~\citep{Phillips2023}, which require no training and reduce the number of iterations needed for convergence compared with classical Jacobi-based schemes. Although not used in this work, the multigrid solver would reduce the computational burden in more complex LSMs. We hope to exploit these advantages for a range of inverse problems in future work.

The NN4PDEs code has recently been used to determine the conductivity of the subsurface given a set of steady-state observations~\citep{Li2024}. Here, we extend Li et~al.'s methodology to a transient application to determine the parameters of a simple land-surface model, and to study whether a parameter can be uniquely determined from the observations (parameter identifiability), whether the effects of different parameters on the model response can be distinguished from one another (confounding) and whether many different parameter combinations result in similar parameter values (equifinality). Although developed independently, our proposed AI-enabled physics-based approach is similar to the NoahPy model which has been developed to improve the representation of permafrost in LSMs~\citep{Tian2026}. In that work, numerical discretisations are formulated as a recurrent neural network, enabling parameter estimation from observations at multiple soil depths. The model estimates four parameters using the Adam optimiser and was shown to converge more efficiently than traditional gradient-free optimisation methods.

The key contribution of this paper is the formulation of a land-surface model within the Neural Physics framework, an end-to-end, fully differentiable, physics-based system that enables efficient parameter estimation. We intentionally chose a simple LSM that still includes a surface energy balance and time-dependence, in order to isolate parameter identifiability, confounding and equifinality. The surface energy balance equation is coupled to a soil layer (Section~\ref{sec:model}), but does not incorporate a subsurface moisture model or explicitly represent buildings. By expressing the governing equations as convolutions within a machine-learning framework using Neural Physics (Section~\ref{sec:discretisation_and_implementation}), we eliminate the need to derive and maintain adjoint models, which is a significant practical barrier in traditional LSM calibration workflows. Instead, we demonstrate that our approach can be applied directly using gradient-based optimisation to estimate parameters in time-dependent problems (Section~\ref{sec:inverse}), a process that can otherwise be tedious~\citep{Lipson2024UP}. 
Using a synthetic dataset for which the correct input parameters are known, the sensitivity to the initial estimate and dependencies between parameters are investigated (Section~\ref{sec:onedepth}). We show that simultaneous measurements at two depths are needed to obtain reliable estimates for the soil properties independent of the initial estimates (Section~\ref{sec:twodepths}). Finally, we use the model to determine the site parameters for a 150-hour period of the flux data recorded in the city of Phoenix, United States and evaluate the model on a subsequent 50-hour period (Section~\ref{sec:Phoenix}). This approach provides insight into the degree of underdetermination in the system (e.g., identifiable parameter combinations and the role of observational configuration), which are often difficult to assess using conventional calibration methods. We comment on this in the results and conclusions (Sections~\ref{sec:results} and~\ref{sec:conclusions}).

\section{Methods} \label{sec:methods}
\subsection{Land-surface model} \label{sec:model}

\begin{figure}[htbp]
\centering
\includegraphics[width=10cm]{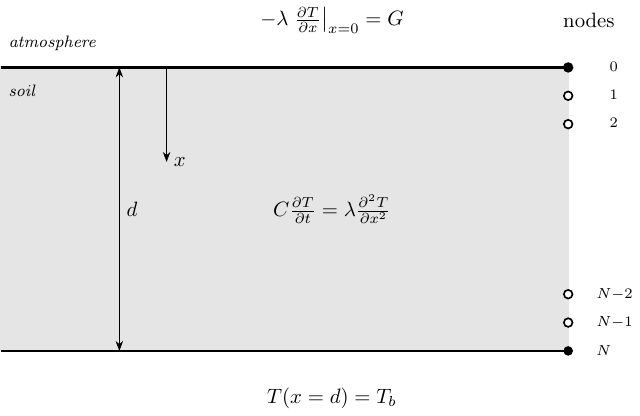}
\small \caption{Definition sketch of the domain with the governing equations and boundary conditions. Shown on the right are the nodes of the discretised problem.} \label{fig:bc}
\end{figure}

The land-surface model considered comprises a soil layer coupled to a surface energy balance at the soil-atmosphere interface (Figure~\ref{fig:bc}). The surface energy balance equation is given by 
\begin{equation}
\begin{split} \label{eq:bc1}
G(t) = & (1-\alpha) K^\downarrow(t) + L^\downarrow(t)-\epsilon \sigma T^4_s(t) 
- h (1+\beta^{-1}) (T_s(t) - T_a(t)),
\end{split}
\end{equation} 
where $G$ is the ground heat flux, $K^\downarrow$ and $L^\downarrow$ are the downwelling shortwave and longwave radiation respectively, $\alpha$ is the surface albedo, $\epsilon$ is the emissivity, $\sigma = 5.67\times 10^{-8}$\unit{W\, m^{-2}\, K^{-4}} is the Stefan-Boltzmann constant, $h$ is the heat transfer coefficient, $T_a$ is the atmospheric temperature and $T_s$ is the absolute surface temperature. The quantity $\beta$ is the Bowen ratio, defined as 
\begin{equation} \label{eq:bowen}
\beta  = \frac{Q_H}{Q_E} ,
\end{equation}
where $Q_H$ is the sensible heat flux and $Q_E$ is the latent heat flux.
The enthalpy equation for the soil is given by 
\begin{equation}\label{eq:diffusion-advection}
 C \frac{\partial T}{\partial t} = \lambda \frac{\partial^2 T}{\partial x^2},
\end{equation}
where $x$ is a downward coordinate starting at the surface, $T(x,t)$ is the soil temperature, $C$ is the volumetric heat capacity and $\lambda$ is the thermal conductivity. The boundary conditions are given by
\begin{equation}
-\lambda\frac{\partial T}{\partial x}(x=0,t) = G(t), \qquad
T(x=d,t) = T_b, \label{eq:bc2}
\end{equation}
where $d$ is the vertical extent of the soil layer and $T_b$ is the fixed temperature at the bottom boundary of the soil layer. The initial condition is given by
\begin{equation} 
  T(x,t=0) = T_b. 
\end{equation}

\subsection{Forward model}\label{sec:discretisation_and_implementation}

\subsubsection{Discretisation}\label{sec:discretisation}
Time and space are discretised uniformly as $t^n = n \Delta t$ and $x_i = i \Delta x$, respectively, where $\Delta t$ is the time increment, $\Delta x = d/N$ is the grid increment and $N$ is the number of intervals. 
Using a central finite difference discretisation in space and a backward Euler discretisation in time, 
\eqref{eq:diffusion-advection} becomes
\begin{equation} \label{eq:2ndor}
    -rT^{n+1}_{i+1}+(1\!+\!2r)T^{n+1}_{i}-rT^{n+1}_{i-1} = T^{n}_{i} \quad \forall i\in\{1,\ldots, N-1\}\,, 
\end{equation}
where $T_i^n = T(x_i, t^n)$ and $r = \lambda\Delta t/({C\Delta x^2})$. The backward Euler scheme is chosen here because it is absolutely stable and allows for large time-steps, which is useful for the computational efficiency of the inverse model. 

The surface boundary condition in~\eqref{eq:bc1} is imposed by choosing a second order central discretisation for the surface flux and combining this with the PDE at the boundary \citep{Bradle}, which after discretisation with an Euler backward time-integration method becomes:
\begin{equation}
\label{eq:bc1_euler}
    \frac{T_0^{n+1}-T_0^{n}}{\Delta t} = \frac{2 \lambda}{C \Delta x} \left(\frac{T_1^{n+1}-T_0^{n+1}}{\Delta x} + \frac{G(t^{n+1})}{\lambda} \right),
\end{equation}
where the ground heat flux $G(t^{n+1})$ uses a mixed discretisation for the outgoing longwave radiation term: 
\begin{equation}\label{eq:bc1_discretised}
G(t^{n+1}) =  (1-\alpha) K^\downarrow(t^{n+1}) + L^\downarrow(t^{n+1})
-\epsilon \sigma  \left(T_0^{n+1}\right)^4 - h (1+\beta^{-1}) \left(T_0^{n+1} - T_a(t^{n+1})\right).
\end{equation}
The discretised bottom boundary condition is given by 
\begin{equation}\label{eq:bc2_discretised}
    T^{n+1}_{N} = T_b\,.
\end{equation}

In order to avoid inverting a matrix, the system (\ref{eq:2ndor}-\ref{eq:bc2_discretised}) is solved using Jacobi iterations~\citep{Greenbaum}. For the PDE discretisation in \eqref{eq:2ndor}, this implies an iteration scheme of the form
\begin{equation}\label{eq:jacobi}
T^{n+1, (k)}_{i} = \frac{T^{n}_{i} + r \left(T^{n+1, (k-1)}_{i+1}+T^{n+1, (k-1)}_{i-1}\right)}{1+2r},
\end{equation}
where $k$ denotes the iteration number.
The surface boundary condition~\eqref{eq:bc1_discretised} is solved by linearising the outgoing longwave radiation term $\left(\text{i.e. }\epsilon \sigma  \left(T_0^{n+1,(k-1)}\right)^3 T_0^{n+1,(k)}\right)$.

\subsubsection{Implementation of Forward Model with Neural Physics}

The NN4PDEs code of Neural Physics uses a convolutional layer with pre-determined weights to implement the finite difference scheme described in Section~\ref{sec:discretisation}. Equation~\eqref{eq:jacobi} can be re-written as
\begin{equation}\label{eq:finite_diff_dics_conv}
T^{n+1, (k)}_{i} = \frac{T^{n}_{i}}{1+2r} +\left.\bm{w}\ast\bm{T}^{n+1,(k-1)}\right|_{i}
\end{equation}
in which $\ast$ represents the discrete convolution, defined as
\begin{equation}\label{eq:definition_dc}
\left.\bm{w}\ast\bm{T}\right|_{i} 
\equiv \sum_{j=-1}^{1} w_{j} T_{i+j},
\end{equation}
where $\bm{T}^{n+1,(k)}$ is a vector containing the nodal values of temperature at iteration~$k$, and the vector $\bm{w}$ has values $[r
/(1 + 2r), 0, r /(1 + 2r)]$. \eqref{eq:finite_diff_dics_conv} and~\eqref{eq:definition_dc} apply to all interior nodes $i \in \{1,\ldots,N-1\}$.
The discretisation in~\eqref{eq:finite_diff_dics_conv} is analogous to a convolutional layer in a neural network where the filter or kernel has weights given by~$\bm{w}$. 

The PyTorch machine-learning library is used to build the forward model. To initialise the neural network, the convolution function and solver for the implicit time-stepping scheme are created. Within the solver, Jacobi iteration for interior nodes is performed by repeated application of the convolution function  \eqref{eq:finite_diff_dics_conv}. In each iteration, the boundary temperatures $T_0^{n+1}$ and $T_N^{n+1}$ are updated and appended to the vector for interior nodes to give the full temperature vector. Once the error, which is taken as the maximum absolute difference between consecutive approximations, is smaller than the defined tolerance, the latest temperature vector is saved and used as the new guess for temperature at the next time level. The iteration procedure is repeated for subsequent time steps until the final time is reached. Convergence checks have been performed to ensure that the numerical scheme is stable and consistent \citep{Huang2024}. 

\begin{table}
\caption{
Input parameter ranges and values for the land-surface model. The "Value" column contains the parameter values used to create the synthetic dataset. Data sourced from \citet{Brutsaert_2005,oke2002boundary,LALOUI202069,Oke2017urban}.
}  
\label{tab:parameters}
\centering 
    \begin{tabular}{p{5cm}llll} \toprule
    \textbf{Variable} & & \textbf{Typical range} & \textbf{Value} & \textbf{Units} \\\toprule
    \multicolumn{4}{l}{\textbf{Assumed Constants}} \\ \midrule
    Emissivity (bare soil) & $\epsilon $ & 0.95--0.98 &0.95&$-$  \\ 
    Total depth & $d$ & 0.5--10.0 &1.0& \unit{m} \\ \midrule
    \multicolumn{5}{l}{\textbf{Time series}} \\  \midrule
    Direct solar irradiance & $S_b$ &$-$ & 800 & \unit{W\, m^{-2}} \\
    Re-emitted longwave radiation & $L^\downarrow$ &$-$& 100 & \unit{W\, m^{-2}}  \\
    Atmospheric temperature & $T_a$ & $-$ & 296 & \unit{K}  \\ \midrule
    \multicolumn{5}{l}{\textbf{Parameters to be determined}} \\ \midrule 
    Albedo & $\alpha $ & 0.05 -- 0.5 & 0.2 & $-$ \\
    Volumetric Heat capacity & $C$ & Sandy soil: 1.28  -- 2.96  &2.2& \unit{MJ\, m^{-3}\, K^{-1}} \\
    & & Clay soil: 1.42 -- 3.10  & & \\
    & &  Peat soil: 0.58  -- 4.02  & & \\ 
    Thermal conductivity & $\lambda$ & Sandy soil: 0.30  -- 2.20  &0.8& \unit{W\, m^{-1}\, K^{-1}} \\
    & &  Clay soil: 0.25 -- 1.58 & & \\ 
    & &  Peat soil: 0.06  -- 0.50  & & \\ 
    Initial / boundary temperature & $T_b$ & Warm: 273 -- 303 &293& \unit{K} \\
    &  & Snow: 263 -- 301 &\\ 
    &  & Arid: 268 -- 308 &\\ 
    Heat transfer coefficient & $h$ & 5 -- 30 &15& $\unit{W\, m^{-2}\, K^{-1}}$ \\
    Bowen Ratio & $\beta$ & 0.5 -- 10 &1.5& $-$ \\\bottomrule
    \end{tabular}
\end{table}

\subsection{Inverse model}\label{sec:inverse}
The inverse problem aims to find a set of parameters $\bm{p}$ that could have given rise to a set of observations. First, an initial guess is made for the values of the parameters. Second, the forward model is solved through time. Third, a data mismatch functional is calculated, based on the difference between the model output and the observations:
\begin{equation}\label{eq:functional}
J(\bm{p}) =  \frac{1}{|\mathcal{I}|\,|\mathcal{N}|}\sum_{i\in\mathcal{I}} \sum_{n\in\mathcal{N}} \left( T(x_i,t^n; \bm{p}) - \tilde{T}(x_i,t^n) \right)^2,
\end{equation}
where $T(x_i,t^n; \bm{p})$ denotes the model prediction for parameters $\bm{p}$, $\tilde T(x_i, t^n)$ denotes the observations, $\mathcal{I}$ is a set of indices which indicate where the observations are taken and $\mathcal{N}$ is a set of indices indicating when there are observations. The cardinality or size of these sets is represented by $|\mathcal{I}|$ and $|\mathcal{N}|$. The final stage is to minimise this mismatch with respect to the parameters:
\begin{equation}
\bm{p^*} = \argmin_{\bm{p}} J(\bm{p})\,,
\end{equation}
where $\bm{p^*}$ is the optimised parameter vector. Classical optimisation methods such as stochastic gradient descent can be used to minimise the data mismatch and calculate an iterative update of the parameters based on the gradient~\citep{Fletcher, Nocedal}.  We use a modified version of stochastic gradient descent known as Adam, which scales the step length with estimates of the mean and variance of the gradients during the iterative process, in order to improve the speed and stability of the algorithm~\citep{Kingma2017}. Each parameter has an associated step length, the initial value of which set according to the order of magnitude of the parameter. $C$ has the largest initial step length of~$10^4$, $T_b$ has an initial step length of~$1$, and the remaining parameters have initial step lengths of~$0.1$. We note that in a machine-learning context, step length is more often referred to as learning rate. However, as we are an using optimisation method to minimise a functional without training a neural network, we use the term step length, in line with classical texts on optimisation~\citep{Fletcher, Nocedal}.

\begin{figure}[htbp]
\centering
\includegraphics[width=17cm]{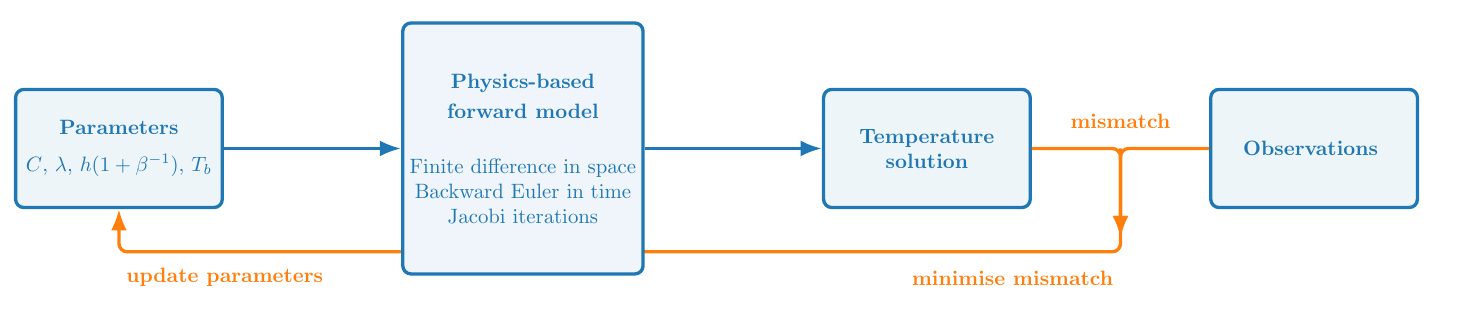}
\caption{\label{fig:framework}A schematic diagram of the inverse problem. Initial guesses for a set of parameters are fed into the forward model (here, a physics-based model expressed as convolutions using Neural Physics) resulting in a solution for temperature (blue arrows). The mismatch between temperature and the observations is minimised with gradient descent and the parameters are updated (orange arrows). This process continues until convergence.}
\end{figure}

The optimisation procedure involves solving the forward model, minimising the mismatch between the model output and the observations using the Adam method to give an improved estimation of parameter values that are consistent with the observations. These steps are repeated until the maximum number of iterations (defined as 150) is reached, or the solution has converged. The solution is considered converged at iteration $\bm{n}$ if both the relative change in the data mismatch functional $J$ is less than $0.01\%$  
and the relative changes in all parameters $\bm{p}$ are less than $0.1\%$. This procedure is illustrated in Figure~\ref{fig:framework}. The optimisation process is expedited using Python's multi-processing library and the coding platform Google Colab. Each active session in Google Colab uses 2~CPUs, allowing a maximum of 10~trials to be run simultaneously with 5~active sessions.
  
\section{Results}
\label{sec:results}
\subsection{Observations at a single depth}\label{sec:onedepth}
To investigate the performance of the inverse model, we produce synthetic data using the forward model, so that the correct parameter values are known.
We initially pursue a `naive' approach, which is to simply run the inverse model to determine the six main input parameters, namely $\bm{p} = \{\alpha$, $\beta$, $h$, $C$, $\lambda$, $T_b\}$, without consideration of the physics of the underlying model. 

Table~\ref{tab:parameters} shows all the inputs that are needed to obtain model output, including the parameter values used to create the synthetic dataset. There is relatively little variation in the emissivity $\epsilon$ of longwave radiation, and we will therefore simply assume a fixed value of 0.95. The soil thickness must be chosen such that it is larger than the penetration thickness $d_\omega = \sqrt{8 \pi^2  \lambda/(C\omega)}$ \citep{Carslaw_Jaeger}, where $\omega$ is the angular frequency of a diurnal cycle. For the synthetic dataset, $d_\omega=0.628\,\unit{m}$, which means that $d=1\,\unit{m}$ is sufficiently deep to assume that the temperature remains unchanged at that depth. 
The soil layer is discretised with $N=100$ intervals. A large number of intervals is used to ensure that spatial discretisation errors will not affect the results.

The shortwave radiation varies according to
\begin{equation}
K^{\downarrow}(t) = S_b \cos(\theta (t)).
\end{equation}
Here, $\theta$ is the solar zenith angle, which is the angle between the direction of the sun's rays and the axis perpendicular to the ground, and $S_b$ is the direct solar irradiance when $\theta$ = 0.
The solar zenith angle $\theta$ is paramaterised as
\begin{equation*}
  \theta (\hat t) = 
  \begin{cases}
    \! 
    \begin{alignedat}{2}
      &-\frac{\pi}{2} & \text{if } \hat t \leqslant 7;
      \\
      &-\frac{\pi}{2} + \frac{\pi}{14} (\hat t - 7) \quad  & \text{if } 7 <\hat t < 21 ;
      \\
      &\frac{\pi}{2} & \text{if } \hat t \geqslant 21 
    \end{alignedat}
  \end{cases}
\end{equation*}
where $\hat t $ represents the hour in the day and is given by $t/3600  \bmod 24$. It is assumed that the sun rises at $7am$, and sets at $9pm$. Before $7am$ and after $9pm$, $\theta =  -\frac{\pi}{2}$ which gives zero $K^{\downarrow}$. Between $7am$ and $9pm$, $K^{\downarrow}$ varies sinusoidally and peaks at $2pm$. The longwave radiation $L^{\downarrow}$ and atmospheric temperature $T_a$ are assumed constant.
The values assumed for $S_b$, $L^{\downarrow}$ and $T_a$ are stated in Table~\ref{tab:parameters}. The simulation is run for 100\,\unit{hours} and the temperatures recorded over time at a depth of 5\,\unit{cm} are used as the observations to determine all six parameters with the inverse model. 

In order to investigate whether the parameters to be determined are sensitive to the initial parameter values, 50 trials are performed, each starting from different initial parameter values. The parameters for trial~$i$ are initialised using  $\bm{p}_{i}=\bm{p}_{i}^{min}+R(\bm{p}_{i}^{max}-\bm{p}_{i}^{min})$, where $\bm{p}_{i}^{min}$ and $\bm{p}_{i}^{max} $ are the minimum and maximum parameter values defined such that the range of the initial values is sufficiently large, but still lies within the typical ranges of the parameter (see Table~\ref{tab:parameters}), and $R$ is a random number taken from a uniform probability distribution.  

\begin{figure*} 
    \centering
    \includegraphics[width=1 \linewidth]{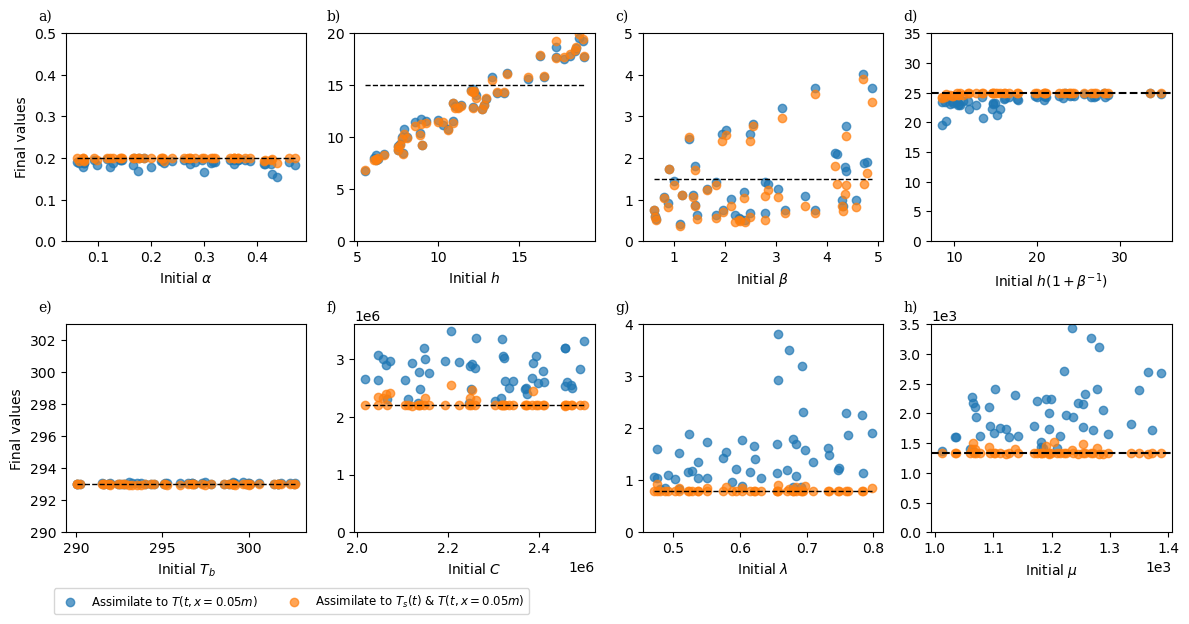}
    \caption{Scatter plots of initial vs final values for the synthetic data set of (a) surface albedo $\alpha$, (b) heat transfer coefficient $h$, (c) Bowen ratio $\beta$, (d) $h(1+\beta^{-1})$, (e) Initial/interior temperature $T_b$, (f) volumetric heat capacity $C$, g) thermal conductivity $\lambda$ and h) $\mu$ compared with their true values (black dotted line). The blue circles assimilate data from depth of 0.05\,\unit{m} and the orange circles assimilate data from two heights, 0.05\,\unit{m} and 0\,\unit{m} (i.e., the surface). 
    }
    \label{fig:cov}
\end{figure*} 

The results from the 50 trials are shown in Figure~\ref{fig:cov} in the blue circles. Scatter plots are shown for each parameter, with its initial value on the $x$-axis and its final value on the $y$-axis. Clearly, the ideal outcome is that the same final value is reached regardless of the initial parameter value, but this is not always the case. The results indicate that the algorithm is able to determine $T_b$ and $\alpha$. However, the value of $h$ is strongly dependent on its initial value value. The parameters $\beta$, $\lambda$ and $C$ have a substantial spread in their optimal values, in a seemingly uncorrelated manner with their initial value.
Hence, the conclusion from the `naive' parameter estimation is that it is not possible to obtain a reliable estimation of the parameter values by assuming the parameters are independent and using a single observation of the soil temperature profile.

\begin{figure} 
    \centering
    \includegraphics[width=0.48\linewidth, trim=0mm 130mm 0mm 0mm, clip]{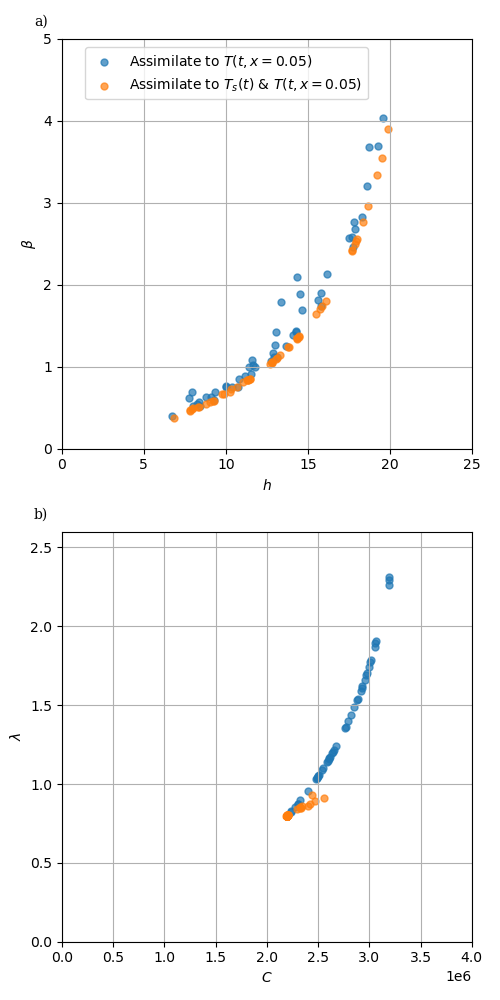}~%
    \raisebox{-3mm}{%
    \includegraphics[width=0.48\linewidth, trim=0mm 0mm 0mm 125mm, clip]{f04_hbeta_Clambda_cov.png}
    }
    \it\caption{Covariance plots of final parameter values for the synthetic dataset of (a) $\beta$ against $h$; and (b) $\lambda$ against $C$}
    \label{fig:hbeta_Clambda_cov}
\end{figure}

The reason why this happened is because an optimiser converges to a global minimum only when the optimisation problem is convex. This condition will be violated when two parameters are dependent on each other. In particular, the heat transfer coefficient $h$ and the Bowen ratio $\beta$ occur only through the product $h(1+\beta^{-1})$ in \eqref{eq:bc1}. As a result, the optimisation problem is non-unique. Figure~\ref{fig:hbeta_Clambda_cov}(a) shows the final values for $h$ and $\beta$ plotted against each other, showing a clear relation between the two quantities. Figure~\ref{fig:cov}(d) shows that the product $h(1+\beta^{-1})$ has much less spread than $h$ and $\beta$ individually. Thus, the optimisation should use the parameter $h(1+\beta^{-1})$ instead of $h$ and $\beta$ separately. 

Furthermore, it is well known that thermal inertia is typically characterised using the thermal admittance $\mu=\lambda C$ \citep{Oke2017urban}, which characterises the ability of a material  to absorb and release heat. This suggests that there might also be a correlation between $C$ and $\lambda$.  Figure~\ref{fig:hbeta_Clambda_cov}(b) shows the final values for $C$ and $\lambda$ plotted against each other, once more showing a clear relation between the two quantities. However, Figure~\ref{fig:cov}(h) shows that even though $C$ and $\lambda$ are clearly correlated, there is considerable uncertainty in $\mu$, and it is not possible to replace $C$ and $\lambda$ for $\mu$.

To conclude this section, we have learnt that a `naive' parameter estimation approach does not result in reliable parameter values. Firstly, the parameters can be correlated to each other, and we found that it is not possible to determine $h$ and $\beta$ individually.
However, even though it is shown that $C$ and $\lambda$ are strongly correlated, the quantity $\mu$ does not capture this behaviour and we thus conclude that it is impossible to obtain reliable parameter values using temperature observations at a single soil depth.

\subsection{Observations at two depths}\label{sec:twodepths}

In this section, parameter estimation using two sets of observations of temperatures is explored, once more using the synthetic data. In addition to $T$($t$; $x$=0.05\,\unit{m}), the surface temperature $T_s$ will be used, since this is where the temperature amplitudes are largest. Instead of changing the parameter estimation strategy and using $h (1+\beta^{-1})$ as a model parameter, we will keep using $h$ and $\beta$, but with the understanding that we should only look whether the quantity $h (1+\beta^{-1})$ has a value that is independent of its initial value.

Figure~\ref{fig:cov} shows that by using observations at two depths, it becomes possible to reliably estimate parameter values. The parameter values obtained from the inverse model are $\alpha=0.199 \pm 0.010$, $T_b=292.99 \pm 0.05\,\unit{K}$, $\lambda=0.81 \pm 0.11\,\unit{W\, m^{-1}\, K^{-1}}$, $C=2.24 \pm 0.32 \,\unit{MJ\, m^{-3}\, K^{-1}}$ and $h(1+\beta^{-1})=24.90 \pm 0.78\,\unit{W\, m^{-2}\, K^{-1}}$. Obtained from 50 trials, the mean values of the parameters are within 2\% of their real value (see Table \ref{tab:parameters}). Note that $C$ and $\lambda$ are the only quantities that have outliers up to 15\% larger than the mean value.

\subsection{Application to Phoenix dataset}\label{sec:Phoenix}

In order to apply the inverse model to real data, we use data from a flux tower managed by the CAP LTER programme. The tower is located in a residential suburban of Maryvale, in the city of Phoenix, United States. The forcing data (incoming shortwave radiation $K^\downarrow$, re-emitted longwave radiation $L^\downarrow$ and atmospheric temperature $T_a$) were collected over the calendar year of 2012 \citep{Chowetal, Chow}. The data was then bias-corrected and gap-filled by the Urban-PLUMBER project~\citep{essd-14-5157-2022,Lipson2024UP}. Phoenix is known for its arid, desert-like climate. This minimises the effect of the moisture content of soil on heat flux exchange, which is not included in the equations of the current model. We extracted 200 hours of flux data from 1 May 2012 0:00~am to 9 May 2012 8:00~am. The first 150 hours of soil temperature measurements are used for data assimilation to estimate the parameters, whilst the remaining 50 hours of data are used for evaluation of the model and its parameters. A longer time series is used than the synthetic data for assimilation to account for the error in assuming a uniform initial temperature profile, which is unlikely to occur in real life. For detailed information about the measurement instruments in the West Phoenix flux tower, see~\ref{app:Phoenix}.

For this dataset, we aim to estimate four parameters, namely $T_b$, $h(1+\beta^{-1})$, $\lambda$ and $C$. The value of $\alpha$ will not be determined using optimisation, as we found that the parameter values could not be determined reliably without specifying $\alpha$, presumably since $\alpha$ controls the absorbed shortwave radiation which is the primary driver of the system. However, this value can be inferred directly from measured incoming and outgoing short-wave radiations. In this case, it is taken to be $\alpha=0.172$, an average mid-day value estimated for the site by \citet{essd-14-5157-2022,Lipson2024UP}. Once more, a total of 50 trials are conducted using initial parameter values that are randomly generated using the procedure discussed in Section~\ref{sec:onedepth}. Table~\ref{tab:final} shows the final parameter values from assimilating to temperature measurements at 5\,\unit{cm} and 15\,\unit{cm}, which are taken to be the mean of the optimised values from all 50 trials.

\begin{table}
\caption{Estimated parameters for the West Phoenix flux tower data set for the period of 1 May 00:00~am -- 9 May 08:00~am from assimilating soil temperatures at 0.05\,\unit{m} and 0.15\,\unit{m}.}
\centering 
    \begin{tabular}{ccccc} \toprule
    Parameter & Units & Mean & Standard deviation \\\toprule
    $C$ & \unit{MJ\, m^{-3}\, K^{-1}} & 2.44 & 0.08 \\
    $\lambda$ & \unit{W\, m^{-1}\, K^{-1}} & 0.826 & 0.043 \\
    $h(1+\beta^{-1})$ & \unit{W\, m^{-2}\, K^{-1}} & 19.66 & 0.11 \\
    $T_b$ & \unit{K} & 305.41 & 0.05\\
     \bottomrule
    \end{tabular}
    \label{tab:final}
\end{table}

Comparing results of inverse analysis using real data to that of synthetic data when assimilated to measurements at two depths (Figure~\ref{fig:cov}), $\lambda$ has a larger distribution in its optimised values, indicating higher uncertainties. The estimated thermal conductivity $\lambda$ and volumetric heat capacity $C$ are within the range of typical values for all three types of soils. The estimated boundary temperature $T_b$ is also reasonable for an arid climate during summer (Table~\ref{tab:parameters}). 

\begin{figure*} 
    \centering
    \includegraphics[width=1 \linewidth]{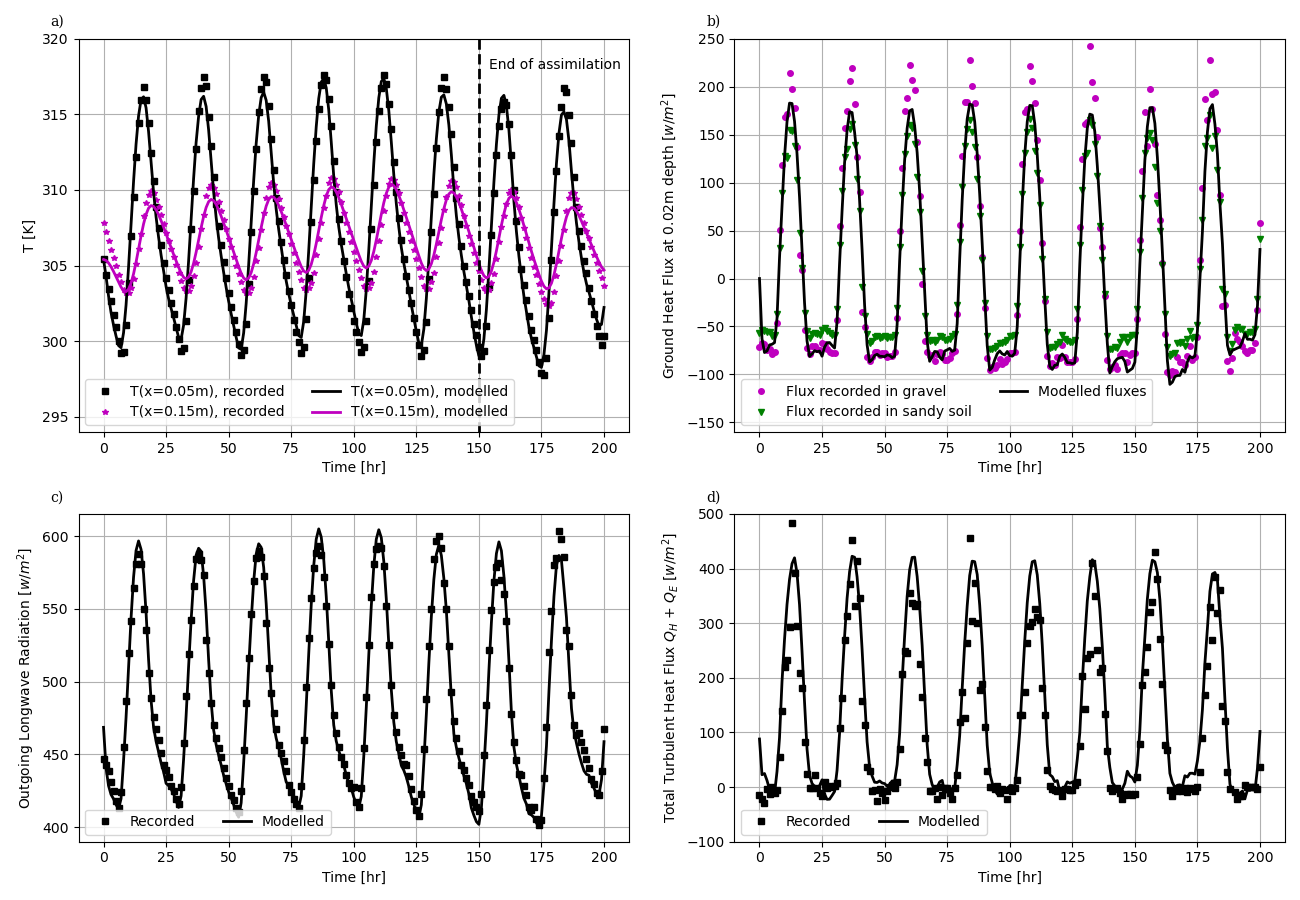}
    \it\caption{Comparison of modelled and recorded data for the West Phoenix flux tower dataset. a) Temperatures at 0.05\,\unit{m} and 0.15\,\unit{m} depth. b) Ground heat flux at 0.02\,\unit{m} depth. c) Outgoing long-wave radiation. d) Combined sensible-latent heat flux}
    \label{fig:finaloutput}
\end{figure*}

In order to compare the modelled fluxes with the observations, the forward model is run using the mean of the obtained parameter values. Figure~\ref{fig:finaloutput}(a) shows good agreement of the model output for the two temperature time series with the final 50 hours of observations, which are not used for the optimisation. More interesting are comparisons with other observed quantities, such as the outgoing longwave radiation $L^\uparrow$, the combined sensible-latent heat flux $Q_E+Q_H$ and conductive soil fluxes in the soil. Figure~\ref{fig:finaloutput}(b) shows the conductive soil flux measurements in  gravel and sandy soil as observed together with the model outputs. For the model, the flux $G=-\lambda \partial T/\partial x$ is estimated using central differences. The modelled ground heat flux lies in between the ground heat flux measured in gravel and sandy soil, as it resembles the flux recorded in sandy soil more in the day, and flux recorded in gravel more at night.  Figure~\ref{fig:finaloutput}(c) compares the observed upwelling longwave radiation $L^\uparrow$ with the model output, once more showing good agreement between the two. It is not self-evident that this should be the case, since the long-wave radiation picks up information from buildings as well as the soil, and the surface temperature has a much larger amplitude than the temperatures deeper in the soil, thus allowing for extrapolation errors. Finally the combined sensible-latent heat fluxes $Q_H+Q_E$ are calculated and compared with flux tower measurements in Figure~\ref{fig:finaloutput}(d). Here, the agreement is not as good as for the other surface energy balance components. We note that the differences between the observations and the model for the turbulent fluxes could be due to the difference in scale of the forcing data and target temperature measurements. Turbulent fluxes are measured at a height of 22.1\,\unit{m} AGL and represent an averaged value across the source area, which is highly variable as it changes with wind direction and atmospheric stability \citep{Chowetal,Schmid91}, whilst soil temperature measurements are localised (taken only at one location beneath the flux tower).

\section{Conclusions}
\label{sec:conclusions}
We developed a land-surface model that enables end-to-end parameter estimation using gradient-based optimisation through Neural Physics, an approach which takes advantage of the tools from machine-learning libraries to provide differentiability whilst remaining entirely physics based. 
The approach also offers insights into parameter identifiability and confounding, and we provide a concrete, quantitative characterisation of equifinality for this application,  identifying specific pathways through which it can be mitigated. 

Results from a `naive' approach that assumes parameter independence show that temperature observations at a single depth are not sufficient to obtain reliable parameter values. In fact, the problem is shown to be ill-posed as multiple parameter combinations (e.g., $h$ and $\beta$) collapse onto an identifiable product $h(1+\beta^{-1})$ and soil parameters ($\lambda$, $C$) exhibit strong correlations with a large spread in their individual estimates. We demonstrate that this non-identifiability is not intrinsic to the model alone, but depends on the observational configuration, as adding temperature observations at a second depth resolves this degeneracy and enables reliable recovery of thermal conductivity and heat capacity, independent of initial conditions. We also show that some confounding cannot be resolved without additional physics or data, for example, the sensible and latent heat fluxes remain inseparable without direct flux measurements. The model is then applied to observations of a 150-hour period of the West Phoenix flux tower dataset and evaluated on a subsequent 50-hour period of data. The estimated parameter values for the thermal conductivity, volumetric heat capacity and the combined sensible-latent heat transfer coefficient are physically plausible, and the agreement of the predictions for outgoing longwave radiation, conductive soil fluxes and combined sensible-latent heat fluxes with observations in the evaluation period is good.

In terms of parameter estimation, we have therefore learnt the following:
\begin{enumerate}
    \item Data at two depths is required in order to obtain reliable estimates for thermal conductivity and heat capacity.
    \item It is not possible to partition sensible and latent heat fluxes without specific measurements of either $Q_H$ or~$Q_E$.
    \item It is necessary to specify the effective albedo of the site, as without it parameter values did not become independent of their initial estimate. We surmise that this is because the effective albedo is the primary forcing during day-time and all fluxes will be affected by this. Since this is one of the simplest quantities to measure, this is not considered a major limitation.
    \item We recommend performing optimisation with at least 50 random initial parameter estimates to ensure the robustness of the obtained parameter values and to take the mean value over all the trials. However, our results suggest that it is possible to use a lower number.  
\end{enumerate}

Several challenges remain for future work. First, the LSM used to demonstrate the proposed approach is intentionally simple. The absence of a subsurface moisture model allows a detailed investigation of parameter identifiability and equifinality, but also motivates future work aimed at extending the framework to more realistic settings. In the present study, we therefore focus on a period from the West Phoenix flux tower dataset during which latent heat fluxes are negligible (i.e., no rainfall). Applying the approach to operational LSMs will require methods capable of estimating the larger number of parameters associated with more complex process representations. In addition, the inverse model is applied over a limited time period, avoiding the need to assume slowly varying parameter values. This assumption becomes particularly important when considering soil moisture, which is inherently time-dependent and typically requires multiple state variables for accurate representation~\citep{Grimmond2010PILPSPhase1}. For longer time series, more sophisticated approaches may be required, potentially including the incorporation of a learned component. Although it is feasible to augment the current model with such methods, the parameter estimation results indicate that the present physics-based formulation is already capable of reproducing observations with reasonable accuracy. 

The second challenge concerns practical adoption of the framework. LSMs are typically implemented in legacy languages such as Fortran, whereas machine-learning frameworks are commonly developed in Python, JAX or Julia. Therefore, applying the Neural Physics approach, or, indeed, adapting code to any framework with automatic differentiability, requires some degree of model refactoring. Although challenging, this process is becoming increasingly tractable with modern tools such as Github Copilot, Claude Code and ChatGPT~\citep{Zhou2024}. 

A third challenge is to deal with the non-smooth parameterisations arising from more complex LSMs. Many operational land-surface models include discontinuities arising from lookup tables, threshold-based parameterisations and conditional logic, which can complicate gradient-based optimisation in practice. Although modern automatic differentiation frameworks can handle piecewise-smooth operations, extending our approach to discontinuous or hybrid formulations is an important direction for future work and an active area of research.

Despite these challenges, the proposed framework offers a potential advantage when scaling to more complex systems. \citet{Dukes2026} identify differentiable hybrid physics--machine-learning models as a new and promising direction for parameter estimation in Earth System Models, but highlight the substantial computational burden associated with increasing model complexity. In particular, stability constraints often necessitate smaller time steps, leading to higher iteration counts and increased memory demands in inverse problems. In this context, the Neural Physics formulation provides a natural way of incorporating fully differentiable multigrid solvers through U-Net  architectures~\citep{Chen2024,Chen2024multiphase,Phillips2023}. Such approaches are well suited to reducing iteration counts, thereby mitigating a key computational bottleneck. Whilst not required for the simple case considered here, this capability represents a potential advantage of the proposed approach when applied to more complex LSMs.

\appendix
\section{West Phoenix flux tower data}
\label{app:Phoenix}
Information about the West Phoenix flux data measurements is provided in Table~\ref{tab:data}.
\begin{table}[htbp]
\caption{Summary of West Phoenix flux data \label{tab:data}}
\begin{threeparttable}
\centering 
    \begin{tabular}{lp{3cm}lp{3cm}p{3cm}lp{1.5cm}} \toprule
    \textbf{Variable} & \textbf{Description} & \textbf{Units} & \textbf{Instrument} & \textbf{Location} & \textbf{Bias correction} & \textbf{Gap filled} \\\toprule
    \multicolumn{7}{p{1\textwidth}}{\textbf{Forcing data\tnote{1}~~\citep{lipson_zenodo, essd-14-5157-2022,Lipson2024UP}}} \\ \midrule
    $K^\downarrow$ & Downward shortwave radiation & \unit{W\, m^{-2}} & Net radiometer/NR01 & 22.1\,\unit{m} AGL & none & 1.63 \% gap-filled \\
    $L^\downarrow$ & Downward longwave radiation & \unit{W\, m^{-2}} & Net radiometer/NR01 & 22.1\,\unit{m} AGL & hourly and daily & 1.32 \% gap-filled \\
    $T_a$ & Air Temperature & \unit{K} & Temperature–relative humidity sensor/HMP45AC & 22.1\,\unit{m} AGL & hourly and daily & 2.95 \% gap-filled \\\midrule
    \multicolumn{7}{p{1\textwidth}}{\textbf{Outgoing Radiation \& Soil temperature measurements \citep{Chowetal, Chow}}} \\ \midrule
    $K^\uparrow$ & Upward shortwave radiation & \unit{W\, m^{-2}} & Net radiometer/NR01 & 22.1\,\unit{m} AGL & none & no \\
    $L^\uparrow$ & Upward longwave radiation & \unit{W\, m^{-2}} & Net radiometer/NR01 & 22.1\,\unit{m} AGL & none & no \\
    $Q_{E}$ & Latent heat flux & \unit{W\, m^{-2}} & Eddy Covariance & 22.1\,\unit{m} AGL & none & no \\
    $Q_{H}$ & Sensible heat flux & \unit{W\, m^{-2}} & Eddy Covariance & 22.1\,\unit{m} AGL & none & no \\
    $-$ & Soil temperature \tnote{2} & \unit{K} & Soil averaging thermocouple/TCAV & 0.05\,\unit{m} \& 0.15\,\unit{m} BGL & none & no \\\bottomrule
    \end{tabular}
\begin{tablenotes}
\item[1] Urban-PLUMBER gap-filled the forcing data using a combination of three methods: contemporaneous and nearby flux tower data where available, linear interpolation for small gaps ($\leqslant 2h$), and bias-corrected ERA5 reanalysis meteorological data for larger gaps. ERA5 combines satellite, atmospheric and ground-based observations to produce globally consistent forcing data at 0.25$^{\circ}$ spatial and hourly temporal resolutions; \item[2] Soil temperature measurements are only available from March 2012
\end{tablenotes}
\end{threeparttable}
\end{table}

\subsection*{Code availability}
The code that supports the findings of this study is available on the Zenodo server under \url{https://doi.org/10.5281/zenodo.19344692} \citep{huang_2026_19344692}. 

\subsection*{Author contribution} 
All authors contributed to the development of methods; RH developed the software, implemented the test cases, analysed the results and prepared the visualisations; RH wrote the paper with contributions from CEH and MvR; CEH and MvR conceptualised and supervised the project.

\subsection*{Competing Interests}
The contact author has declared that none of the authors has any competing interests.

\subsection*{Acknowledgements}
The authors would like to thank Prof.\ Sue Grimmond for the insightful conversations about atmospheric temperature readings.
CEH\ gratefully acknowledges support from Imperial-X's Eric and Wendy Schmidt Centre for AI in Science and funding from the EPSRC AI-Respire project (grant number EP/Y018680/1). MvR acknowledges funding from the NERC ASSURE project (grant number NE/W002868/1).
The authors would also like to thank the reviewers, whose valuable suggestions have helped improve the manuscript.

\bibliographystyle{plainnat}


\end{document}